\documentclass[12pt]{article}
\pdfoutput=1
\usepackage{draft,tikz-cd,multirow}

\usepackage{tikz,tikz-3dplot}
\usepackage{pgfplots}
\pgfplotsset{compat=1.17}
\usetikzlibrary{shapes,arrows,cd,chains,decorations.markings,decorations.pathmorphing,calc}
\tikzset{
	->-/.style args={#1rotate#2}{decoration={markings, mark=at position #1 with {\arrow[scale=1.5,rotate = #2 ]{stealth}}}, postaction={decorate}}
}
\tikzstyle{GraphNode}=[circle, draw=black, fill=black, inner sep=2pt, minimum size=5pt]
\tikzstyle{GraphEdge}=[black]
\pgfmathsetmacro{\gS}{1}

\begin{document}

\begin{titlepage}

\begin{center}

\title{Witten index of BMN matrix quantum mechanics}

\author{Chi-Ming Chang}

\address{Yau Mathematical Sciences Center (YMSC), Tsinghua University, Beijing 100084, China}

% \address{School of Natural Sciences, Institute for Advanced Study, Princeton, NJ 08540, USA}

\address{Beijing Institute of Mathematical Sciences and Applications (BIMSA) \\ Beijing 101408, China}

\email{cmchang@tsinghua.edu.cn}

\end{center}

\vfill

\begin{abstract}
We compute the Witten index of the Berenstein-Maldacena-Nastase matrix quantum mechanics, which counts the number of ground states as well as the difference between the numbers of bosonic and fermionic BPS states with nonzero angular momenta. The Witten index sets a lower bound on the entropy, which exhibits $N^2$ growth that provides strong evidence for the existence of BPS black holes in M-theory, asymptotic to the plane-wave geometry. We also discuss a relation between the Witten index in the infinite $N$ limit and the superconformal index of the Aharony-Bergman-Jafferis-Maldacena theory.

\end{abstract}

\vfill

\end{titlepage}

\tableofcontents

\section{Introduction}

The Berenstein-Maldacena-Nastase (BMN) matrix quantum mechanics \cite{Berenstein:2002jq} is a supersymmetric gauged quantum mechanics with sixteen supercharges. It is a mass deformation of the Banks-Fischler-Shenker-Susskind (BFSS) matrix quantum mechanics \cite{Banks:1996vh}, and converges to the BFSS matrix quantum mechanics when the dimensionless effective gauge coupling $g$ (normalized by the mass scale) approaches infinity.

The BMN matrix quantum mechanics participates in a very rich array of dualities. First, by the BFSS conjecture \cite{Banks:1996vh}, the BMN matrix quantum mechanics is dual to M-theory on the uncompactified plane-wave background \cite{Kowalski-Glikman:1984qtj}, when the rank $N$ of the matrices approaches infinity with the gauge coupling $g$ fixed. A stronger version of the conjecture states that, at finite $N$ and $g$, the BMN matrix quantum mechanics is dual to the Discrete Light-Cone Quantization (DLCQ) of M-theory on the plane wave background with $N$ units of light-cone momentum \cite{Susskind:1997cw,Sen:1997we,Seiberg:1997ad}. Finally, in the 't Hooft limit ($\lambda=g^2 N$ fixed and $N\to\infty$), the BMN matrix quantum mechanics is dual to M-theory on asymptotically plane-wave backgrounds \cite{Itzhaki:1998dd,Polchinski:1999br}, for instance, at large 't Hooft coupling $(\lambda\gg 1)$, the vacuum states are dual to the Lin-Lunin-Maldacena geometries \cite{Lin:2004nb,Lin:2005nh}, and the deconfined phase at high temperature is dual to black holes \cite{Costa:2014wya}.

Witten index is a very powerful tool in the study of supersymmetric quantum mechanics \cite{Witten:1982df}. It is defined as a trace over the Hilbert space, weighted by the Boltzmann factor and an additional insertion of the fermion parity operator. The trace is equal to the number of bosonic ground states minus the number of fermionic ground states because there are equal numbers of bosonic and fermionic states at each non-zero energy level of the Hamiltonian due to supersymmetry. Hence, the Witten index is independent of the temperature and any deformation of the Hamiltonian as long as the deformation is gentle enough that it does not change the Hilbert space. When the theory has global symmetries that commute with the Hamiltonian, the Witten index can be further refined by incorporating flavor fugacities and thus receives contributions not only from ground states but also from BPS states with non-zero symmetry charges.

The Witten index of the BMN matrix quantum mechanics can be computed in the weak coupling (large mass) limit due to protection by supersymmetry. In this limit, the classical vacua of the theory are separated by infinitely high potential barriers, so excitations around each vacuum define mutually decoupled superselection sectors. Each sector comprises free bosonic and fermionic harmonic oscillators \cite{Dasgupta:2002hx}, and its spectrum is obtained by acting with an arbitrary numbers of creation operators on the ground state. These ingredients assemble into a matrix-integral formula for the Witten index (see Section~\ref{sec:computation}), closely analogous to the superconformal index of ${\cal N}=4$ super-Yang–Mills (SYM) \cite{Kinney:2005ej}. This result differs from the Witten index of the BFSS matrix quantum mechanics \cite{Yi:1997eg,Sethi:1997pa}: in the BFSS (infinite-coupling, zero-mass) limit, the flat directions in the classical potential modify the Hilbert space relative to that of the finite-coupling theory.

The Witten index restricted to different superselection sectors exhibits very different behavior in the large $N$ limit. We will focus on the sectors associated with the trivial vacuum  (the maximally reducible vacuum) and the irreducible vacuum. In the trivial vacuum sector, the matrix integral can be evaluated directly as a power series in the fugacities at finite $N$, or using a saddle-point approximation in the large $N$ and small chemical potentials limit. Both results exhibit $N^2$ growth of the entropy. This strongly suggests the existence of BPS black holes, analogous to the non-BPS black holes \cite{Costa:2014wya}, in the bulk dual of the BMN matrix quantum mechanics (see Section~\ref{sec:index_trivial_vacuum}). In the irreducible vacuum sector, the BMN matrix quantum mechanics describes a single M2-brane with $N$ units of light-cone momentum along the M-theory circle \cite{Berenstein:2002jq}. On the other hand, a system of coinciding M2-branes is described by the Aharony-Bergman-Jafferis-Maldacena (ABJM) theory \cite{Aharony:2008ug}. We find that the Witten index in the irreducible vacuum sector in the large $N$ limit is equal to the superconformal index of the ${\rm U}(1)_{1}\times {\rm U}(1)_{-1}$ ABJM theory \cite{Kim:2009wb} up to a divergent factor (see Section~\ref{sec:index_irreducible_sector}).

\section{BMN index}
\label{eqn:BMN_index}
\subsection{BMN matrix quantum mechanics}

The Hamiltonian of the BMN matrix quantum mechanics is \cite{Maldacena:2002rb}
\ie\label{eqn:BMN_Hamiltonian}
&H=R\,\Tr\left[\frac12 \sum_{A=1}^9(P^A)^2-\frac 1{4\ell_P^6}\sum^9_{A,B=1}[X^A,X^B]^2-\frac1{2\ell_P^3}\Psi^T\gamma^A[X^A,\Psi]\right]
\\
&\hspace{-.4cm}+\frac R2\,\Tr\left[\left(\frac\mu{3R}\right)^2\sum^3_{i=1}(X^i)^2+\left(\frac\mu{6R}\right)^2\sum^9_{a=4}(X^a)^2+i\frac\mu{4R}\Psi^T\gamma^{123}\Psi+i\frac{2\mu}{3R\ell_P^3}\epsilon_{ijk}X^i X^j X^k\right]\,,
\fe
where $X^A$ for $A=1,\,\cdots,\,9$ are nine $N\times N$ bosonic matrices, $\Psi$ with a spinor index suppressed denotes sixteen $N\times N$ fermionic matrices, and $P^A$ are the conjugate momenta of $X^A$. $X^i$ and $X^a$ for $i=1,\,2,\,3$ and $a=4,\,\cdots,\,9$ are the first three and last six components of $X^A$. We will focus on the ${\rm U}(N)$ gauge group, and the gauge symmetry acts on the Hermitian matrices  $X^A$, $\Psi$, and $P^A$ via conjugation by unitary matrices. The parameters $R$, $\ell_P$ and $1/\mu$ have dimensions of length. $\ell_P$ can be absorbed by field and parameter redefinitions: $X^A = \ell_P \widetilde X^A$ and $R = \ell_P^2 \widetilde R$, so that $\widetilde X^A$ are dimensionless and $1/\widetilde R$ has dimension of length. Therefore, there is only a single dimensionless coupling 
\ie
g^2=\frac{R^3}{\mu^3 \ell_P^6}\,.
\fe

At infinite coupling $g\to\infty$ ($\mu=0$), the BMN matrix quantum mechanics becomes the Banks-Fischler-Shenker-Susskind (BFSS) matrix quantum mechanics \cite{Banks:1996vh}, with the Hamiltonian given by the first line of \eqref{eqn:BMN_Hamiltonian}, which preserves ${\rm SO}(9)$ symmetry. $X^A$, $P^A$ are in the vector representation and $\Psi$ is in the spinor representation. The potential terms on the second line of \eqref{eqn:BMN_Hamiltonian} break ${\rm SO}(9)$ to ${\rm SO}(3)\times {\rm SO}(6)$. 

\subsection{Witten index}

The BMN matrix quantum mechanics has ${\rm SU}(2|4)$ supersymmetry, which contains the ${\rm SO}(3)\times {\rm SO}(6)$ symmetry as the maximal bosonic subgroup. The ${\rm SU}(2|4)$ supersymmetry contains sixteen supercharges $Q^I_\alpha$ and $(Q^I_\alpha)^\dagger$, where $\alpha=\pm$ is the spinor index of ${\rm SU}(2)\cong {\rm SO}(3)$, and $I=1,\cdots,4$ is the index for the spinor representation of ${\rm SO}(6)$ (fundamental representation of ${\rm SU}(4)$). The supercharges have the anti-commutator \cite{Dasgupta:2002hx}
\ie\label{eqn:BMNQQ}
\{Q^I_\alpha,(Q^J_{\beta})^\dagger\}=2\delta^I_J \delta^\alpha_\beta H+\frac{\mu}{3}\epsilon_{ijk}(\sigma^k)_\A^\B\delta^I_J M^{ij}-\frac{2\mu}{3}\delta^\A_\B R^I_J\,,
\fe
where $M^{ij}$ and $R^I_J$ are the rotation generators of ${\rm SO}(3)$ and ${\rm SU}(4)$, respectively. Let us pick a supercharge $Q:= Q^4_-$. The anti-commutator of $Q$ with its Hermitian conjugate $Q^\dagger$ is
\ie\label{eqn:QQdagger}
2\Delta:=\{Q,Q^\dagger\} =2H-\frac{2\mu}{3} M^{12}-\frac{2\mu}{3} R^4_4=2H-\frac{2\mu}{3} M^{12}-\frac{\mu}{3} (M^{45}+M^{67}+M^{89})\,,
\fe
where we expand $R^4_4$ in terms of the Cartan generators $M^{45}$, $M^{67}$, $M^{89}$ of ${\rm SO}(6)$, which correspond to the rotations along the three orthogonal two-planes in ${\mathbb R}^6$. It is convenient to assemble the 5 Cartan generators of ${\rm SU}(2|4)$ into a vector as
\ie\label{eqn:charge_vector}
\left(\frac{12H}{\mu},4M^{12},2M^{45},2M^{67},2M^{89}\right)\,.
\fe
The supercharge $Q$ has the Cartan charges $(1,-2,1,1,1)$.

Let us consider the thermal partition function\footnote{In the path-integral formalism, the partition function \eqref{eqn:ther_par} is obtained from the Euclidean action built from the twisted Hamiltonian
\ie
H+ \frac{6\omega- 2\beta\mu}{3\beta}M^{12}+\frac{3\Delta_1- \beta\mu}{3\beta}M^{45}+\frac{3\Delta_2- \beta\mu}{3\beta}M^{67}+\frac{3\Delta_3- \beta\mu}{3\beta}M^{89}\,,
\fe
with imaginary (Euclidean) time compactified as $\tau\sim \tau+\beta$.}
\ie\label{eqn:ther_par}
Z= \Tr\Omega\,,\quad \Omega := e^{-\beta \Delta -2 \omega M^{12}-\Delta_1 M^{45}-\Delta_2 M^{67}-\Delta_3 M^{89}}\,.
\fe
The Boltzmann factor $\Omega$ anti-commutes with the supercharge $Q$ if the chemical potentials satisfy the linear relation
\ie\label{eqn:chem_rel}
\Delta_1+\Delta_2+\Delta_3 -2\omega = 2\pi i\mod 4\pi i\,.
\fe
The Witten index is defined by
\ie\label{eqn:PWMM_index}
{\cal I}=Z\big|_{\eqref{eqn:chem_rel}}=\Tr \left[(-1)^{2M^{12}} e^{-\beta\Delta-\Delta_1 (M^{12}+M^{45})-\Delta_2 (M^{12}+M^{67})-\Delta_3 (M^{12}+M^{89})}\right]\,,
\fe
where the factor $(-1)^{2M^{12}}$ follows directly from imposing \eqref{eqn:chem_rel}. In the block decomposition used below, this insertion can differ from the ordinary fermion parity on individual bi-fundamental letters by a block-dependent sign, but these signs cancel in gauge-invariant words. The Witten index ${\cal I}$ is independent of the inverse temperature $\beta$, because the states with nonzero $\Delta$-eigenvalues are all paired up by the action of the supercharge $Q$ and their contributions to the Witten index cancel. Since all the explicit coupling constant $g$ dependence is inside the Hamiltonian, the Witten index is also independent of $g$ away from the infinite coupling ($g=\infty$) point where the dimension of the Hilbert space becomes uncountable due to continua in the spectrum.

\subsection{Computation at weak coupling}
\label{sec:computation}
Since the Witten index is independent of the coupling constant $g$, it can be computed for the BMN matrix quantum mechanics in the weak coupling (small $g$) limit, or equivalently, in the large $\mu$ limit. We will begin by reviewing the computation of the weak coupling spectrum in \cite{Dasgupta:2002hx}, and then use the results to compute the Witten index.

In the large $\mu$ limit, the bosonic potential becomes very steep, and we can expand the Hamiltonian about the minima of the potential (classical supersymmetric vacua). The bosonic potential can be written in a manifestly positive form as
\ie
V=\frac R2\Tr\left[\left(\frac\mu{3R}X^i+\ell_P^{-3}i\epsilon^{ijk}X^j X^k\right)^2+\frac1{2\ell_P^2}(i[X^a,X^b])^2+\frac1{\ell_P^2}(i[X^a,X^i])^2+\left(\frac\mu{6R}\right)^2(X^a)^2\right]\,.
\fe
The classical supersymmetric vacua are given by \cite{Berenstein:2002jq}
\ie
X^a=0\,,\quad X^i=\frac{\mu \ell_P^3}{3R}J^i\,,
\fe
where $J^i$ is a $N$-dimensional matrix representation of ${\rm SU}(2)$, i.e. $[J^i,J^j]=i\epsilon^{ijk}J^k$. Any $N$-dimensional representation of ${\rm SU}(2)$ can be written as a direct sum of irreducible representations, and is labeled by a partition of the integer $N$,
\ie\label{eqn:partition_N}
N=\sum_{k=1}^K n_k N_k\,.
\fe
More explicitly, the matrix $J^i$ decomposes into $K\times K$ blocks as
\ie
J^i=\sum_{k,l=1}^KJ^i_{kl}\,,\quad J^i_{kl}=\delta_{kl}I_{n_l}\otimes J^i_l \,,
\fe
where $J^i_l$ is the $N_l$-dimensional irreducible representation, and $I_{n_l}$ is an $n_l\times n_l$ identity matrix. In the trivial vacuum ($K=1$, $n_1=N$, and $N_1=1$), the ${\rm U}(N)$ gauge symmetry is preserved. In the nontrivial vacua, the ${\rm U}(N)$ gauge symmetry is broken to
\ie\label{eqn:unbroken_subgroup}
{\rm U}(n_1)\times\cdots\times{\rm U}(n_K)\,.
\fe
The classical vacua correspond to degenerate ground states, which are protected quantum mechanically and have vanishing energy at finite coupling $g>0$ \cite{Kim:2002if,Dasgupta:2002ru,Kim:2002zg}.

The expansion about the classical vacua was studied in \cite{Dasgupta:2002hx}. In the small $g$ limit, the Hamiltonian reduces to a free quadratic term plus higher order interaction terms, which are suppressed by powers of $g$ relative to the quadratic term. Hence, the theory divides into superselection sectors associated with each classical vacuum and labeled by the partition \eqref{eqn:partition_N} of $N$. Each sector contains a single ground state, and excited states given by acting creation operators of a set of bosonic and fermionic harmonic oscillators. More explicitly, according to the integer partition \eqref{eqn:partition_N}, we decompose the matrices $X^A$ and $\Psi$ into $K\times K$ blocks, and further write the $(k,l)$-th block as a sum of tensor products of an $n_k\times n_l$ matrix and an $N_k\times N_l$ matrix. For example, the matrix $X^a$ has the expansion
\ie
X^a=\sum^K_{k,l=1}X^a_{kl}\,,\quad X^a_{kl}&=\sum_{j=\frac12 |N_k-N_l|}^{\frac12 (N_k+N_l)-1}\sum^{j}_{m=-j}x^a_{kl,jm}\otimes Y^{N_kN_l}_{jm}\,,
\fe
where $X^a_{kl}$ is the $(k,l)$-block of $X^a$. $Y^{N_kN_l}_{jm}$ is an $N_k\times N_l$ matrix that is a spin-$j$ representation in the tensor product of the spin-$\left(\frac{N_k-1}{2}\right)$ and spin-$\left(\frac{N_l-1}{2}\right)$ representations. $x^a_{kl,jm}$ is an $n_k\times n_l$ matrix whose components are bosonic harmonic oscillators. Similar decompositions of $X^i$ and $\Psi$ give the bosonic harmonic oscillators
\ie
\alpha_{kl,jm}\,,\quad \beta_{kl,jm}\,, 
\fe
and the fermionic oscillators
\ie
\chi^I_{kl,jm}\,,\quad \eta_{I,kl,jm}\,, 
\fe
all of which are $n_k\times n_l$ matrices and transform in the bi-fundamental representation of ${\rm U}(n_k)\times {\rm U}(n_l)$.\footnote{Note that the matrix indices of the $n_k\times n_l$ matrices are implicit.} The ranges of $j$ for each oscillator are given in Tables 1 and 2 in \cite{Dasgupta:2002hx}.

We focus on BPS excited states, which are given by acting with creation operators of the BPS letters, the oscillators satisfying the BPS condition $\Delta=0$. The allowed values of $j=M^{12}$ differ by a half-unit between the bosonic and fermionic BPS letters: when $N_k-N_l$ is even, the bosonic letters have integer $M^{12}$ while the fermionic letters have half-integer $M^{12}$; when $N_k-N_l$ is odd, this parity assignment is reversed. Thus, on an individual $(k,l)$ BPS letter, one has $(-1)^{2M^{12}}=(-1)^F(-1)^{N_k-N_l}$, so $(-1)^{2M^{12}}$ agrees with the ordinary fermion parity only for even $N_k-N_l$. The extra block-dependent sign cancels in gauge-invariant words. The BPS letters are listed in Table~\ref{tab:BPS_letters}.

\begin{table}[H]
\begin{center}
\begin{tabular}{|c|c|c|}
\hline
letter & charges & index
\\\hline
$\beta $ & $(4j,4j,0,0,0)$ & $e^{-2j\omega}$
\\\hline
\multirow{3}{*}{$x$}   & $(4j+2,4j,2,0,0)$ &  $e^{-2j\omega-\Delta_1}$
\\\cline{2-3}
   & $(4j+2,4j,0,2,0)$ &  $e^{-2j\omega-\Delta_2}$
\\\cline{2-3}
  & $(4j+2,4j,0,0,2)$ &  $e^{-2j\omega-\Delta_3}$
\\\hline
$\chi$ & $(4j+3,4j,1,1,1)$ & $e^{-2j\omega -\frac{\Delta_1}{2}-\frac{\Delta_2}{2}-\frac{\Delta_3}{2}}$
\\\hline
\multirow{3}{*}{$\eta$} & $(4j+1,4j,1,1,-1)$ & $e^{-2j\omega -\frac{\Delta_1}{2}-\frac{\Delta_2}{2}+\frac{\Delta_3}{2}}$
\\\cline{2-3}
& $(4j+1,4j,1,-1,1)$ & $e^{-2j\omega -\frac{\Delta_1}{2}+\frac{\Delta_2}{2}-\frac{\Delta_3}{2}}$
\\\cline{2-3}
& $(4j+1,4j,-1,1,1)$ & $e^{-2j\omega +\frac{\Delta_1}{2}-\frac{\Delta_2}{2}-\frac{\Delta_3}{2}}$
\\\hline
\end{tabular}
\end{center}
\caption{\label{tab:BPS_letters}The BPS letters and their charges (in the vector form \eqref{eqn:charge_vector}) and indices.}
\end{table}

For the $(k,l)$-th BPS letters, the bosonic and fermionic single-letter partition functions are
\ie
Z^{\rm single,B}_{kl}(\omega,\Delta_i,U_k,U_l)
&=z^{\rm B}_{kl}(\omega,\Delta_i)\chi_{\text{bi-fund}}(U_k,U_l)\,,
\\
Z^{\rm single,F}_{kl}(\omega,\Delta_i,U_k,U_l)
&=z^{\rm F}_{kl}(\omega,\Delta_i)\chi_{\text{bi-fund}}(U_k,U_l)\,.
\fe
Here the fugacities $\omega$ and $\Delta_i$ are defined previously in \eqref{eqn:ther_par}, $U_k$ is a $n_k\times n_k$ unitary matrix, and $\chi_{\text{bi-fund}}$ is the bi-fundamental character
\ie
\chi_{\text{bi-fund}}(U_k,U_l)=\Tr(U_k^\dagger)\Tr(U_l)\,.
\fe
The colorless bosonic and fermionic single-letter partition functions are
\ie
z^{\rm B}_{kl}(\omega,\Delta_i)&=
\sum_{j=\frac12 |N_k-N_l|}^{\frac12 (N_k+N_l)-1} e^{-2 j \omega}(e^{-\Delta_1}+e^{-\Delta_2}+e^{-\Delta_3})
+\sum_{j=\frac12 |N_k-N_l|+1}^{\frac12 (N_k+N_l)} e^{-2 j \omega}\,,
\\
z^{\rm F}_{kl}(\omega,\Delta_i)&=
\sum_{j=\frac12 |N_k-N_l|-\frac12}^{\frac12 (N_k+N_l)-\frac32} e^{-2 j \omega-\frac{\Delta_1}{2}-\frac{\Delta_2}{2}-\frac{\Delta_3}{2}}
-\delta_{N_k,N_l}e^{\omega-\frac{\Delta_1}{2}-\frac{\Delta_2}{2}-\frac{\Delta_3}{2}}
\\
&\quad +\sum_{j=\frac12 |N_k-N_l|+\frac12}^{\frac12 (N_k+N_l)-\frac12}e^{-2 j \omega}
(e^{-\frac{\Delta_1}{2}-\frac{\Delta_2}{2}+\frac{\Delta_3}{2}}+e^{-\frac{\Delta_1}{2}+\frac{\Delta_2}{2}-\frac{\Delta_3}{2}}+e^{\frac{\Delta_1}{2}-\frac{\Delta_2}{2}-\frac{\Delta_3}{2}})\,.
\fe
Before imposing the relation \eqref{eqn:chem_rel} among the chemical potentials, the free multi-letter partition function in the superselection sector specified by the integer partition \eqref{eqn:partition_N} is obtained by taking the plethystic exponential of the bosonic and fermionic single-letter partition functions and then projecting to gauge singlets under the unbroken subgroup \eqref{eqn:unbroken_subgroup}. This gives
\ie\label{eqn:free_partition_integral}
Z^{\rm BMN,free}_{n_i;N_i}(\omega,\Delta_i)
&=\int \prod_{k=1}^K [dU_k]\,
\exp\Biggl[
\sum_{m=1}^\infty\sum_{k,l=1}^K\frac{1}{m}
\\
&\qquad\quad \times
\left(Z^{\rm single,B}_{kl}+(-1)^{m+1}Z^{\rm single,F}_{kl}\right)
(m\omega,m\Delta_i,U_k^m,U_l^m)
\Biggr]\,.
\fe
The factor $(-1)^{m+1}$ is the standard fermionic sign in the plethystic exponential. The Witten index is then obtained from this free multi-letter partition function by imposing \eqref{eqn:chem_rel}. Equivalently, the $m$-th single-letter index\footnote{In a previous version of this manuscript, this index was obtained by imposing \eqref{eqn:chem_rel} on the single-letter partition function before taking the plethystic exponential. This missed the $m$-dependence of the sign in the $m$-th plethystic term. We thank Eunwoo Lee for pointing out this error. This correction does not affect the subsequent discussion: all explicit evaluations below are in one-block sectors, for which $N_k=N_l$ and the corrected sign agrees with the previous expression.} entering the index matrix integral is
\ie\label{eqn:single_letter_index}
\iota^{(m)}_{kl}(\Delta_i)&=\left[z^{\rm B}_{kl}(m\omega,m\Delta_i)+(-1)^{m+1}z^{\rm F}_{kl}(m\omega,m\Delta_i)\right]\big|_{\eqref{eqn:chem_rel}}
\\
&=\sum_{j=\frac12 |N_k-N_l|}^{\frac12 (N_k+N_l)-1}
(-1)^{2jm+1} e^{-jm(\Delta_1+\Delta_2+\Delta_3)}
(1-e^{-m\Delta_1})(1-e^{-m\Delta_2})(1-e^{-m\Delta_3})
+\delta_{N_k,N_l}\,.
\fe
Thus the Witten index in each superselection sector is
\ie\label{eqn:WI_as_U_integral}
{\cal I}^{\rm BMN}_{n_i;N_i}&=
Z^{\rm BMN,free}_{n_i;N_i}(\omega,\Delta_i)\big|_{\eqref{eqn:chem_rel}}
\\
&=\int \prod_{k=1}^K [dU_k]\,
\exp\left[\sum_{m=1}^\infty\sum_{k,l=1}^K\frac{1}{m}\iota^{(m)}_{kl}(\Delta_i)\Tr (U^{\dagger m}_k)\Tr(U_l^m)\right]\,.
\fe
The total Witten index is given by summing over the contributions from all the superselection sectors
\ie
{\cal I}^{\rm BMN}=\sum_{n_i,N_i\in\bZ_{>0},\, \eqref{eqn:partition_N}}{\cal I}^{\rm BMN}_{n_i;N_i}
\fe

\subsection{Trivial vacuum sector}
\label{sec:index_trivial_vacuum}
Let us consider the Witten index in the trivial vacuum sector given by $K=1$, $n_1=N$, $N_1=1$. The integral formula \eqref{eqn:WI_as_U_integral} reduces to
\ie\label{eqn:index_trivial_vacuum}
\hspace{-.34cm}{\cal I}^{\rm BMN}_{N;1}=\int [dU] \exp\left\{\sum_{m=1}^\infty\frac{1-(1-e^{-m\Delta_1})(1-e^{-m\Delta_2})(1-e^{-m\Delta_3})}{m}\Tr(U^{\dagger m})\Tr (U^m)\right\}\,.
\fe

The same index ${\cal I}^{\rm BMN}_{N;1}$ was studied in \cite{Choi:2023znd,Choi:2023vdm} but intepreted as the superconformal index of the ${\cal N}=4$ SYM truncated to the BMN sector, which contains only fundamental fields invariant under the chiral ${\rm SU}(2)_R$ rotation. It was argued in \cite{Choi:2023vdm}, following a similar analysis in \cite{Copetti:2020dil,Choi:2021lbk}, that in the large $N$ limit with small and fixed $\Delta_i$, the matrix integral \eqref{eqn:index_trivial_vacuum} has a saddle-point with the eigenvalue distribution
\ie\label{eqn:small_BH_saddle}
\rho(\alpha)=\frac3{4\pi^3}(\pi^2-\alpha^2)\quad{\rm for}\quad \alpha\in(-\pi,\pi)\,.
\fe
The saddle point contributes to the index as
\ie\label{eqn:logZ_BMN}
\log{\cal I}^{\rm BMN}_{N;1}=-\frac{3N^2}{2\pi^2}\Delta_1\Delta_2\Delta_3\,.
\fe
Let us consider the Legendre transform of $\log{\cal I}^{\rm BMN}_{N;1}$ given by the extremization
\ie\label{eqn:entropy_BMN}
&S^{\rm BMN}_{N;1}(M^{12},M^{45},M^{67},M^{89})
\\
&=\underset{\Delta_i,\omega}{\rm ext} \left[\log{\cal I}^{\rm BMN}_{N;1}+2\omega M^{12}+\Delta_1 M^{45}+\Delta_2 M^{67}+\Delta_3 M^{89}\big|_{\eqref{eqn:chem_rel}}\right]\,,
\\
&=\pm 2\pi\sqrt{\frac{2(M^{12}+M^{45})(M^{12}+M^{67})(M^{12}+M^{89})}{3N^2}}-2\pi i M^{12}\,.
\fe
We follow the analysis in \cite{Choi:2018hmj} for black holes in ${\rm AdS}_5$ by setting the imaginary part to zero and selecting the solution with the positive real part, and obtain
\ie
S^{\rm BMN}_{N;1}(M^{12},M^{45},M^{67},M^{89})=2\pi\sqrt{\frac{2M^{45}M^{67}M^{89}}{3N^2}}\,,\quad M^{12}=0\,.
\fe
The result is valid in the large $N$ limit with $\epsilon_i:=M^{2i+2,2i+3}/N^2$ (for $i=1,\,2,\,3$) held fixed and small $\epsilon_i\ll 1$. The function $S^{\text{BMN}}_{N;1}$ sets a lower bound on the entropy of the BPS states; however, for simplicity, we will also refer to $S^{\text{BMN}}_{N;1}$ as the entropy. We find that the entropy $S^{\rm BMN}_{N;1}\sim\sqrt{\epsilon_1\epsilon_2\epsilon_3}N^2$ exhibits $N^2$ scaling, which implies that the theory is in a deconfined phase and may correspond to a BPS black hole in the bulk dual.

A very subtle point of the above analysis is that the contribution from the non-trivial saddle \eqref{eqn:small_BH_saddle} is negative \eqref{eqn:logZ_BMN}, which is smaller than the contribution from the trivial saddle, corresponding to a confined phase, with a constant eigenvalue distribution
\ie
\rho(\alpha)=\frac{1}{2\pi}\,.
\fe
However, the entropy \eqref{eqn:entropy_BMN} contributed by this saddle is of order $N^2$, much larger than the entropy from the trivial saddle. Hence, the deconfined phase is subdominant to the confined phase in the grand canonical ensemble, but dominant in the microcanonical ensemble. The deconfined phase has a negative specific heat $C$ (or, more precisely, the susceptibility) given by\footnote{I thank Sunjin Choi for a discussion on this point.}
\ie
C=\Delta^2\frac{d^2\log {\cal I}^{\rm BMN}_{N;1}}{d\Delta^2}=-\frac{9N^2}{\pi^2}\Delta^3<0\,,
\fe
where we set the chemical potentials to $\Delta:=\Delta_1=\Delta_2=\Delta_3$. These features are similar to the small black holes in ${\rm AdS}_5$ \cite{Choi:2021lbk}, which we review in Appendix~\ref{sec:small_BH}.\footnote{The ``small" black holes in ${\rm AdS}_5$ still exhibit a macroscopic horizon (with size $\gg$ the string scale) and an $N^2$-scaling entropy.}

\begin{figure}[H]
    \centering
\includegraphics[width=0.495\textwidth]{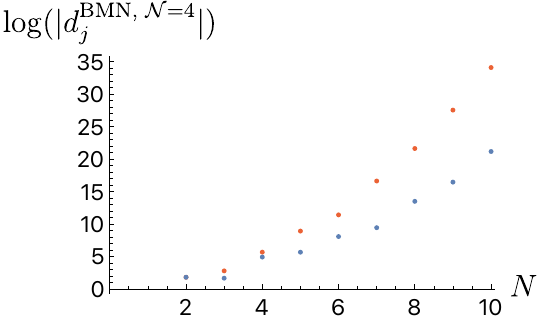}
\includegraphics[width=0.495\textwidth]{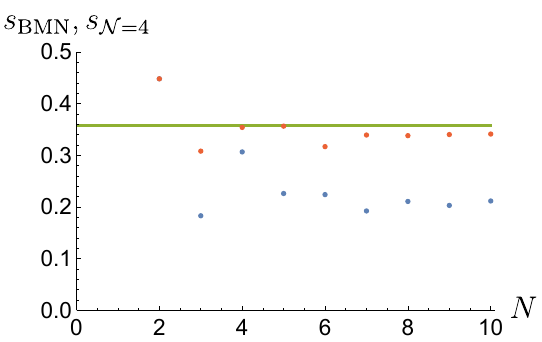}
    \caption{The $\log(|d_j^{\rm BMN}|)$, $s_{\rm BMN}$ (blue) and $\log(|d_j^{{\cal N}=4}|)$, $s_{{\cal N}=4}$ (orange)  at $j=N^2$ with increasing $N$. The green line is the asymptotic value for $s_{{\cal N}=4}$.}
    \label{fig:BMN_numerics}
\end{figure}

An important question is whether the $N^2$ scaling of the entropy persists at finite $\epsilon_i$, and if so, whether the deconfined phase also dominates in the grand canonical ensemble. We provide positive evidence for the $N^2$ scaling in the entropy from evaluating the BMN index \eqref{eqn:index_trivial_vacuum} as an expansion of the fugacities to high powers. In the microcanonical ensemble, the entropy is defined as the log of the coefficients in the expansion of the index
\ie
S_{\rm BMN}(j)=\log |d^{\rm BMN}_j|\,,\quad{\cal I}^{\rm BMN}_{N;1}=\sum_j d^{\rm BMN}_j t^j\,,\quad t^2=e^{-\Delta_1}=e^{-\Delta_2}=e^{-\Delta_3}\,,
\fe
where the quantum number $j$ is the following linear combination of angular momenta
\ie
j=6M^{12}+2(M^{45}+M^{67}+M^{89})%=6J+2(q_1+q_2+q_3)
\,.
\fe
For inspecting the growth of the degeneracy $d^{\rm BMN}_j$ in the large $N$ limit with $j/N^2$ fixed, we define
\ie
s_{\rm BMN}\left(j/N^2\right)=N^{-2}\log |d^{\rm BMN}_j|   \,.
\fe
Let us set $j=N^2$,\footnote{There is a small subtlety that the states in the BMN sector all have even $j$. Therefore, when $N$ is odd, we consider the average $\frac12(\log|d_{j-1}^{\rm BMN}|+\log|d_{j+1}^{\rm BMN}|)$ for $j=N^2$.} and plot $\log|d_j^{\rm BMN}|$ and $s_{\rm BMN}$ against $N$ in Figure~\ref{fig:BMN_numerics}. The BMN index appears to converge when $N\gtrsim6$,
as 
\ie
\log|d_j^{\rm BMN}|\sim 0.21 \times N^2\,.
\fe

For comparison, we perform the same analysis for the superconformal index of the ${\cal N}=4$ SYM, which is given by the matrix integral \cite{Kinney:2005ej}
\ie\label{eqn:N=4_index}
{\cal I}_{{\cal N}=4}=\int [dU] \exp\left\{\sum_{m=1}^\infty\frac{1-\frac{(1-e^{-m\Delta_1})(1-e^{-m\Delta_2})(1-e^{-m\Delta_3})}{(1-e^{-m\omega_1})(1-e^{-m\omega_2})}}{m}\Tr U^{\dagger m}\Tr U^m\right\}\,,
\fe
with the constraint $\Delta_1+\Delta_2+\Delta_3-\omega_1-\omega_2=2\pi i$. We expand the index as
\ie
{\cal I}_{{\cal N}=4}=\sum_j d^{{\cal N}=4}_jt^j\,,\quad t^2=e^{-\Delta_1}=e^{-\Delta_2}=e^{-\Delta_3}\,,~ t^3=e^{-\omega_1}=e^{-\omega_2}\,.
\fe
Again, we consider the large $N$ limit with $j/N^2$ fixed and define
\ie
s_{{\cal N}=4}(j/N^2)=N^{-2}\log |d^{{\cal N}=4}_j|\,.
\fe
The entropy $s_{{\cal N}=4}(j/N^2)$ can be computed exactly in the grand canonical ensemble by a saddle point approximation \cite{Cabo-Bizet:2018ehj,Choi:2018hmj,Benini:2018ywd}. When $j=N^2$, we have $s_{{\cal N}=4}(1)=0.357$. The plots of $\log|d_j^{{\cal N}=4}|$ and $s_{{\cal N}=4}$ against $N^2$ are given in Figure~\ref{fig:BMN_numerics}. We can see that $s_{{\cal N}=4}$ converges to the asymptotic value already at $N\sim 5$.

\subsection{Irreducible vacuum sector}
\label{sec:index_irreducible_sector}

Let us consider the Witten index in the sector given by $K=1$, $n_1=1$, $N_1=N$ corresponding to an irreducible representation of ${\rm SU}(2)$. The integral formula \eqref{eqn:WI_as_U_integral} reduces to
\ie\label{eqn:WI_N=N1}
{\cal I}^{\rm BMN}_{1;N}= \exp\left\{\sum_{m=1}^\infty\frac{1}{m}\left[1-\frac{(1-e^{-m\Delta_1})(1-e^{-m\Delta_2})(1-e^{-m\Delta_3})(1- e^{-mN(\Delta_1+\Delta_2+\Delta_3)})}{1-e^{-m(\Delta_1+\Delta_2+\Delta_3)}}\right]\right\}\,.
\fe
In this sector, the BMN matrix model describes a single M2-brane carrying $N$ units of light-cone momentum along the M-theory circle and wrapping an ${\rm S}^2$ in the plane-wave geometry \cite{Berenstein:2002jq}. On the other hand, the ABJM theory with ${\rm U}(1)_1\times{\rm U}(1)_{-1}$ gauge group in radial quantization also describes a single M2-brane wrapping an ${\rm S}^2$. Hence, we expect some relations between the BMN index and the superconformal index of the ABJM theory \cite{Kim:2009wb}.

In Appendix~\ref{app:ABJM_index}, we review the ${\cal N}=8$ ABJM index. It depends on five chemical potentials associated with one spin and four R-charges of the superconformal symmetry ${\rm OSp}(4|8)$, and with one linear relation between the chemical potentials. However, only the subgroup ${\rm SU}(2|4)\subset {\rm OSp}(4|8)$ is manifest in the BMN matrix quantum mechanics, whose Witten index depends on four chemical potentials with one linear relation. Hence, to compare the BMN index and the ABJM index, one needs to first figure out a relation between the chemical potentials.

\subsubsection*{Relation between the chemical potentials}
Under the embedding ${\rm SU}(2|4)\hookrightarrow {\rm OSp}(8|4)$, the supercharges in ${\rm SU}(2|4)$ are related to the supercharges in ${\rm OSp}(8|4)$ by
\ie
Q^I_\A = \sqrt\frac{\mu}{3}({\bf Q}^{2I-1}_\A+i{\bf Q}^{2I}_\A)\,,\quad (Q^I_\A)^\dagger = \sqrt\frac{\mu}{3}({\bf S}^{2I-1,\A}-i{\bf S}^{2I,\A})\,,
\fe
where $I=1,\,\cdots,\,4$ and ${\bf Q}^{\cal I}_\A$ and ${\bf S}^{{\cal I}\A}$ are the supercharges and the conformal supercharges in ${\rm OSp}(8|4)$. Plugging this into \eqref{eqn:osp84_{Q,S}}, we find the anti-commutator
\ie
\{Q^m_\A,(Q^n_\B)^\dagger\}
=2\delta^m_n\delta^\B_\A H-\frac{2\mu}{3}\delta^m_n J^\B_\A-\frac{2\mu}{3}\delta^\A_\B R^m_n\,,
\fe
where $R^m_n$ are the generators of ${\rm SU}(4)\subset {\rm SO}(8)$. This precisely matches the anti-commutator \eqref{eqn:BMNQQ} in ${\rm SU}(2|4)$, with the identification $J^\B_\A=-\frac12\epsilon_{ijk}(\sigma^i)^\B_\A M^{jk}$. The Cartan generators in ${\rm SU}(2|4)$ and ${\rm OSp}(8|4)$ are related by
\ie\label{eqn:Cartan_su24_to_osp48}
\frac{3}{\mu}H&=D-\frac14({\bf M}^{12}+{\bf M}^{34}+{\bf M}^{56}+{\bf M}^{78})\,,&
M^{12}&=J^-_-\,,
\\
R^1_1&=\frac{1}{4}(3{\bf M}^{12}-{\bf M}^{34}-{\bf M}^{56}-{\bf M}^{78})\,,&
R^2_2&=\frac{1}{4}(-{\bf M}^{12}+3{\bf M}^{34}-{\bf M}^{56}-{\bf M}^{78})\,,
\\
R^3_3&=\frac{1}{4}(-{\bf M}^{12}-{\bf M}^{34}+3{\bf M}^{56}-{\bf M}^{78})\,,&
R^4_4&=\frac{1}{4}(-{\bf M}^{12}-{\bf M}^{34}-{\bf M}^{56}+3{\bf M}^{78})\,,
\fe
where ${\bf M}^{IJ}$ are the ${\rm SO}(8)$ generators.
The Cartan generators $M^{45}$, $M^{67}$, $M^{89}$ of ${\rm SO}(6)\cong {\rm SU}(4)$ are related to the Cartan generators of ${\rm SO}(8)$ by
\ie\label{Cartan_so6_to_so8}
M^{45}&=\frac12({\bf M}^{12}-{\bf M}^{34}-{\bf M}^{56}+{\bf M}^{78})\,,
\\
M^{67}&=\frac12(-{\bf M}^{12}+{\bf M}^{34}-{\bf M}^{56}+{\bf M}^{78})\,,
\\
M^{89}&=\frac12(-{\bf M}^{12}-{\bf M}^{34}+{\bf M}^{56}+{\bf M}^{78})\,.
\fe
Substituting the relations \eqref{eqn:Cartan_su24_to_osp48} and \eqref{Cartan_so6_to_so8} into the formula \eqref{eqn:PWMM_index} of the Witten index, we find the following relation between the fugacities in \eqref{eqn:PWMM_index} and \eqref{eqn:N=8_SCI},
\ie\label{eqn:fugacity_relation}
\gamma_1&=\frac{\Delta_1-\Delta_2-\Delta_3}{2}\,,&\gamma_2&=\frac{-\Delta_1+\Delta_2-\Delta_3}{2}\,,
\\
\gamma_3&=\frac{-\Delta_1-\Delta_2+\Delta_3}{2}\,,&\gamma_4&=\frac{\Delta_1+\Delta_2+\Delta_3}{2}\,.
\fe

\subsubsection*{Matching the indices in the infinite $N$ limit}

Now, we are ready to compare the BMN index and the ABJM index. The ${\rm U}(1)_1\times{\rm U}(1)_{-1}$ ABJM index is given by the integral,
\ie
{\cal I}^{\rm ABJM}_{1,1}&=\sum_{n,\tilde n\in\bZ}y_3^{-\frac12 n}x^{|n-\tilde n|}\int_{-\pi}^\pi \frac{d\alpha d\tilde\alpha}{(2\pi)^2}e^{i(n\alpha-\tilde n\tilde\alpha)}
\\
&\qquad\times\exp\left[\sum_{s=\pm}\sum_{m=1}^\infty \frac1 m x^{m|n-\tilde n|} f_s(x^m,y_1^m,y_2^m)e^{-sm i(\alpha-\tilde\alpha)}\right]\,,
\fe
where the functions $f_\pm(x,y_1,y_2)$ are given in \eqref{eqn:fpm_in_ABJM_index}. The fugacities $x$, $y_1$, $y_2$, $y_3$ are defined in \eqref{eqn:xy_to_gamma}. The integral and sum can be easily performed after the change of the integration variables,
\ie
\alpha_-=\alpha-\tilde\alpha\,,\quad \alpha_+=\frac{\alpha+\tilde\alpha}{2}\,.
\fe
The result is
\ie
{\cal I}^{\rm ABJM}_{1,1}=\exp\bigg[\sum_{m=1}^\infty\frac{1}{m}\frac{e^{-m\gamma_4}}{e^{-2m\gamma_4}-1}\sum_{\substack{s_i=\pm\\s_1 s_2 s_3 s_4=-1}}s_4e^{-\frac m2 (s_1\gamma_1+s_2\gamma_2+s_3\gamma_3+s_4\gamma_4)}\bigg]\,,
\fe
Substituting the relation \eqref{eqn:fugacity_relation} between the fugacities into the above formula, we find
\ie\label{eqn:ABJM11}
 {\cal I}^{\rm ABJM}_{1,1}\big|_{\eqref{eqn:fugacity_relation}}=\exp\left\{\sum_{m=1}^\infty\frac{1}{m}\left[2-\frac{(1-e^{-m\Delta_1})(1-e^{-m\Delta_2})(1-e^{-m\Delta_3})}{1-e^{-m(\Delta_1+\Delta_2+\Delta_3)}}\right]\right\}\,.
\fe
Now, let us compare the ABJM index \eqref{eqn:ABJM11} with the BMN index \eqref{eqn:WI_N=N1} in the $N\to\infty$ limit by taking their ratio
\ie
\frac{{\cal I}^{\rm ABJM}_{1,1}\big|_{\eqref{eqn:fugacity_relation}}}{{\cal I}^{\rm BMN}_{1;\infty}}=\exp\left(\sum_{m=1}^\infty\frac{1}{m}\right)=\infty\,.
\fe
We see that the ABJM index agrees with the BMN index up to a divergent factor.

This divergence in the ABJM index is due to the specialization of the chemical potentials \eqref{eqn:fugacity_relation} from four $\gamma_i$'s to three $\Delta_i$'s. A potential explanation of the divergent factor in the relation between the BMN and ABJM indices is that the M2-brane theory on the sphere has infinitely many degenerate vacua that are not distinguished by the three chemical potentials, $\Delta_i$. However, the BMN quantum mechanics only includes one of these vacua that is chosen by the value of $N$. Hence, in this sense the infinity is due to the sum over $N$, which in the ABJM theory would correspond to tracing over different states with the corresponding chemical potential being set to zero.\footnote{I thank Juan Maldacena for suggesting this explanation.}

The full ABJM index, with all four chemical potentials $\gamma_i$ turned on, could potentially be recovered by performing a sum over the BMN indices with different $N$ weighted by an additional chemical potential. We leave this for future work.

\section{Discussion}

In this paper, we compute the Witten index for the BMN matrix quantum mechanics, which counts (with signs) the number of BPS states. The computation relies on the non-renormalization of the Witten index and is performed at weak coupling, where the theory is divided into superselection sectors associated with each supersymmetric vacuum. The result is a sum of contributions from all superselection sectors, with each term being a matrix integral. In the trivial vacuum sector, we evaluate the matrix integral and show that it contributes $N^2$ growth to the entropy. In the irreducible vacuum sector, we find a novel relation between the BMN index and the ABJM index.

Let us discuss some open problems/future directions:
\begin{itemize}
\item The $N^2$ growth of the entropy in the trivial vacuum sector strongly suggests that there exist BPS black holes in M-theory, asymptotic to the plane-wave geometry. However, no solution of this sort in eleven-dimensional supergravity is currently known. Different BPS black hole geometries, with different horizon topologies, could potentially dominate in different regions of the chemical potentials. It is important to extend the saddle-point analysis of the Witten index in the trivial vacuum sector to finite $\epsilon_i$ and to other vacuum sectors, as it could shed light on the phase structure of these BPS black holes. The phase structure of non-BPS black holes at finite temperature was explored in \cite{Catterall:2010gf,Asano:2018nol,Schaich:2020ubh,Bergner:2021goh,Schaich:2022duk,Pateloudis:2022oos,Pateloudis:2022ijr} on the matrix quantum mechanics side by lattice simulations, and in \cite{Costa:2014wya} on the gravity side.

\item The relation between the BMN and ABJM indices can be extended to the case of multiple M2-branes. On the BMN side, this involves the vacua corresponding to direct sums of multiple copies of $N$-dimensional irreducible ${\rm SU}(2)$ representations. Understanding this generalization may require finding a precise relation between the configurations of M2-branes in these two setups, in particular, the relation between the $N$ of the BMN quantum mechanics and a ${\rm U}(1)$ charge in the ABJM theory.

\item Supersymmetric localization of the BMN matrix quantum mechanics was developed in \cite{Asano:2012zt}, and used in the study of the $1/4$-BPS sector \cite{Asano:2014vba,Asano:2014eca} and the transverse M5-branes \cite{Asano:2017xiy,Asano:2017nxw}. The same techniques may be used to obtain a path integral derivation of the matrix integral formulae \eqref{eqn:WI_as_U_integral} together with \eqref{eqn:single_letter_index} for the Witten index.

\item Following \cite{Kinney:2005ej,Grant:2008sk,Chang:2013fba} and more recently \cite{Chang:2022mjp,Choi:2022caq,Choi:2023znd,Budzik:2023xbr,Choi:2023vdm,Chang:2024zqi}, we can study the $Q$-cohomology of the BMN matrix quantum mechanics.\footnote{The supercharge $Q$ is defined above \eqref{eqn:QQdagger}.} The $Q$-cohomology classes correspond one-to-one with the BPS states, and the Euler characteristic of the $Q$-cohomology equals the Witten index. The $Q$-cohomology can be straightforwardly computed at weak coupling, although there is no proof or argument that the $Q$-cohomology for BMN matrix quantum mechanics remains invariant when varying the coupling constant $g$.

\end{itemize}

\section*{Acknowledgments}

I am grateful to Sunjin Choi, Shota Komatsu, Ying-Hsuan Lin, and Jorge E. Santos for inspiring discussions, and especially to Juan Maldacena for introducing me to this problem and generously sharing many of his important insights.
CC is partly supported by the National Key R\&D Program of China (NO. 2020YFA0713000). 
This research was supported in part by grant NSF PHY-2309135 to the Kavli Institute for Theoretical Physics (KITP). We thank the hospitality of the KITP Program ``What is String Theory? Weaving Perspectives Together'' and Conference ``Spacetime and String Theory".

\appendix

\section{Small black holes in AdS$_5$}
\label{sec:small_BH}

\begin{figure}[htb]
    \centering
\includegraphics[width=0.45\textwidth]{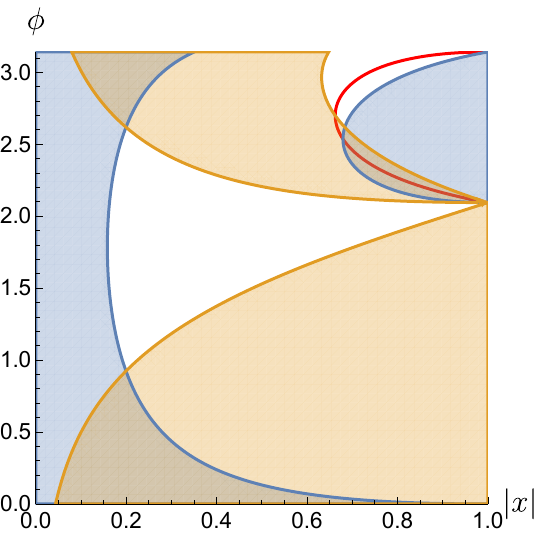}
    \caption{The blue and orange regions are given by \eqref{eqn:phase_bdry} and $C>0$, with $C$ given by \eqref{eqn:specific_heat}. The red curve is given by the function $\beta(Q)$ for $Q$ varying from 0 to $\infty$.}
    \label{fig:phase}
\end{figure}

In this appendix, we review the discussions of small AdS black holes in \cite{Choi:2021lbk}. The BPS black hole in ${\rm AdS}_5$ has a partition function given by the Euclidean gravity path integral:
\ie
\log Z = \frac{N^2}{2}\frac{\Delta_1\Delta_2\Delta_3}{\omega_1\omega_2}\,,\quad \Delta_1+\Delta_2+\Delta_3-\omega_1-\omega_2=2\pi i\,.
\fe
Let us consider the specialization to a single fugacity $x=e^{-\beta}$ as
\ie\label{eqn:chem_spec}
\Delta_i=2\beta+2\pi i\,,\quad \omega_i = 3\beta+2 \pi i\,.
\fe
Since the thermal AdS phase has a free energy of order 1, the black hole phase is in the region
\ie\label{eqn:phase_bdry}
{\rm Re}\,\left(\log Z\right)={\rm Re}\,\left(\frac{N^2}{2}\frac{(2\beta+2\pi i)^3}{(3\beta+2 \pi i)^2}\right) > 0\,.
\fe
We plot this region in blue on the $|x|$--$\phi$ plane in Figure~\ref{fig:phase}, where $x=e^{-\beta}=|x|e^{i\phi}$. 

The large/small black holes lie in regions where the specific heat $C$ (susceptibility) is positive/negative, given by
\ie\label{eqn:specific_heat}
C={\rm Re}\,\left(\beta_{\rm R}^2\frac{d^2\log Z}{d\beta_{\rm R}^2}\right)\,,
\fe
where $\beta_{\rm R}={\rm Re}\,\beta=-\log |x|$. We plot the region of positive specific heat in orange in Figure~\ref{fig:phase}.

The entropy of the black hole is given by extremizing the function
\ie
S(\beta;Q,J)&=\log Z+Q_1\Delta_1+Q_2\Delta_2+Q_3\Delta_3+J_1\omega_1+J_2\omega_2\big|_{Q_i=Q,\,J_i=J,\,\eqref{eqn:chem_spec}}
\\
&=\frac{N^2}{2}\frac{(2\beta+2\pi i)^3}{(3\beta+2 \pi i)^2}+3 Q(2\beta+2\pi i)+2 J(3\beta+2 \pi i)\,.
\fe
This function has three extrema. We pick the one that gives a positive real part of $S$. Imposing the condition that the imaginary part of $S$ vanishes, we find
\ie\label{eqn:relation_QJ}
{\rm Im}\,S=0\quad\Rightarrow \quad Q^3+\frac{N^2}{2}J^2=\left(\frac{N^2}{2}+3Q\right)(3Q^2-N^2 J)\,,
\fe
which admits one solution with $Q,J>0$. The entropy at the chosen extremum is
\ie\label{eqn:entropy_S(Q)}
S(Q)=2\pi\sqrt{3Q^2-N^2 J}\,,
\fe
which is always positive for $Q,J>0$ satisfying the condition \eqref{eqn:relation_QJ}.

Let us write the chosen extremum as $\beta(Q)$, after eliminating/substituting for $J$ via \eqref{eqn:relation_QJ}. We plot $\beta(Q)$ on the $|x|$-$\phi$ plane as the red curve in Figure~\ref{fig:phase} for $Q$ varying from 0 to $\infty$. The small charge limit $Q\to 0$ is at $(|x|,\phi)=(1,\pi)$, and the large charge limit $Q\to \infty$ is at $(|x|,\phi)=(1, \frac{2\pi}{3})$. There is an intersection between the curve $\beta(Q)$ and the phase boundary at
\ie
Q_*=\frac{N^2}{24}(1+\sqrt{33})\,.
\fe
We see that when increasing $Q$ from 0 to $\infty$, the red curve is first inside the white region ($\log Z<0$ and $C<0$) corresponding to a small black hole phase that is subdominant to thermal AdS in the grand canonical ensemble. Then, the red curve enters the orange region ($\log Z<0$ and $C>0$) corresponding to a large black hole phase that is also subdominant. Finally, the red curve enters the orange and blue regions ($\log Z>0$ and $C>0$) corresponding to a large and dominant black hole phase. While the small black holes are always subdominant, they still carry macroscopic entropy \eqref{eqn:entropy_S(Q)}. 

\section{Review of ABJM index}
\label{app:ABJM_index}

\subsection{${\cal N}=8$ superconformal index}
The ${\cal N}=8$ ABJM theory has superconformal symmetry ${\rm OSp}(8|4)$. Let us denote the supercharges and the conformal supercharges of ${\rm OSp}(8|4)$ by
\ie
{\bf Q}^{\mathcal I}_\A\,,\quad {\bf S}^{{\mathcal I}\A}\,,
\fe
where ${\mathcal I}=1,\,\cdots,\,8$ is the vector index of ${\rm SO}(8)$ and $\alpha=\pm$ is the spinor index of ${\rm SU}(2)\subset{\rm Sp}(4)$. We have the anti-commutator
\ie\label{eqn:osp84_{Q,S}}
\{{\bf Q}_\A^{\mathcal I},{\bf S}^{{\mathcal J}\B}\}=\delta^{\mathcal{IJ}}\delta^\B_\A D-\delta^{\mathcal{IJ}} J^\A_\B-i\delta^\A_\B {\bf M}^{\mathcal{IJ}}\,,
\fe
where $D$ is the dilatation charge, $J^\A_\B$ are the generators of ${\rm SU}(2)$, and ${\bf M}^{\mathcal{IJ}}$ are the generators of ${\rm SO}(8)$.
Let us define
\ie
{\bf \Delta}:=\frac12\{{\bf Q}_-^7+i{\bf Q}_-^8,{\bf S}^{7-}-i{\bf S}^{8-}\}=D-J-{\bf M}^{78}\,,
\fe
where $J:= J^-_-$. Let us consider the partition function
\ie
Z=\Tr \Omega\,,\quad \Omega:=e^{-\beta {\bf \Delta}-2\omega J -\gamma_1{\bf M}^{12}-\gamma_2{\bf M}^{34}-\gamma_3{\bf M}^{56}-\gamma_4{\bf M}^{78}}\,.
\fe
The supercharge ${\bf Q}_-^7+i{\bf Q}_-^8$ anti-commutes with the Boltzmann factor $\Omega$ if
\ie\label{eqn:ABJM_fugacity_condition}
\gamma_4-\omega=2\pi i\mod 4\pi i\,.
\fe
We define the superconformal index as
\ie\label{eqn:N=8_SCI}
{\cal I}=Z\big|_{\eqref{eqn:ABJM_fugacity_condition}}=\Tr\left[ (-1)^F e^{-\beta {\bf \Delta}-\gamma_1{\bf M}^{12}-\gamma_2{\bf M}^{34}-\gamma_3{\bf M}^{56}-\gamma_4 (2J+{\bf M}^{78}) }\right]\,,
\fe
where we identify $2J$ with the fermion number $F$, since all the fermionic states have half-integer spin $J$. By the standard arguments, the superconformal index ${\cal I}$ is independent of the inverse temperature $\beta$.

\subsection{Integral formula}
The superconformal index for the ${\rm U}(N)_k\times {\rm U}(N)_{-k}$ ABJM theory was computed in \cite{Kim:2009wb},\footnote{In a flux sector labeled by $\{n_1,\cdots,n_N\}$, the ${\rm U}(N)$ gauge group is broken to $G_{\{n_i\}}={\rm U}(N_1)\times {\rm U}(N_2)\times \cdots$. We have the formula
\ie
\sum_{\{n_i\}}\frac{1}{|G_{\{n_i\}}|}=\frac1{N!}\sum_{n_i\in\bZ}\,,
\fe
where $|G_{\{n_i\}}|$ is the order of the Weyl group of $G_{\{n_i\}}$.}\footnote{Note that our formula differs from the formula (3.11) in \cite{Kim:2009wb} by the replacement $y_3\to y_3^{-1}$.}
\ie\label{eqn:ABJM_index_f1}
{\cal I}^{\rm ABJM}_{N,k}&=\frac1{(N!)^2}\sum_{n_i,\tilde n_i\in\bZ}y_3^{-\frac k2 \sum_{i=1}^Nn_i}\int_{-\pi}^\pi \prod_{i=1}^N\frac{d\alpha_i d\tilde\alpha_i}{(2\pi)^2}e^{ik(n_i\alpha_i-\tilde n_i\tilde\alpha_i)}
\\
&\quad\times\exp\left[-\gamma_4 \epsilon_0+\sum_{m=1}^\infty \frac1 m f(x^m,y_1^m,y_2^m,e^{im\alpha_i},e^{im\tilde\alpha_i})\right]\,,
\fe
where the fugacities $x$ and $y_i$ are
\ie\label{eqn:xy_to_gamma}
x=e^{-\gamma_4}\,,\quad y_1=e^{-\gamma_1}\,,\quad y_2=e^{-\gamma_2}\,,\quad y_3=e^{-\gamma_3}\,.
\fe
The quantities $\epsilon_0$ and $f$ are
\ie
\epsilon_0=\sum_{i,j}|n_i-\tilde n_j|-\sum_{i<j}(|n_i-n_j|+|\tilde n_i-\tilde n_j|)\,,
\fe
and
\ie\label{eqn:fpm_in_ABJM_index}
f(x,y_1,y_2,e^{i\alpha_i},e^{i\tilde\alpha_i})&=-\sum_{i\neq j}x^{|n_i-n_j|}e^{-i(\alpha_i-\alpha_j)}+\sum_{i,j}f_+(x,y_1,y_2)x^{|n_i-\tilde n_j|}e^{-i(\alpha_i-\tilde\alpha_j)}
\\
&\quad-\sum_{i\neq j}x^{|\tilde n_i-\tilde n_j|}e^{-i(\tilde \alpha_i-\tilde\alpha_j)}+\sum_{i,j}f_-(x,y_1,y_2)x^{|\tilde n_i- n_j|}e^{-i(\tilde\alpha_i-\alpha_j)}\,,
\\
f_+(x,y_1,y_2)&=\frac{1}{1-x^2}\left[x^\frac12\left(\sqrt{\frac{y_1}{y_2}}+\sqrt{\frac{y_2}{y_1}}\right)-x^\frac32\left(\sqrt{y_1y_2}+\frac{1}{\sqrt{y_1y_2}}\right)\right]\,,
\\
f_-(x,y_1,y_2)&=f_+(x,y_1,y_2^{-1})\,.
\fe

\bibliography{refs}

\providecommand{\href}[2]{#2}\begingroup\raggedright\begin{thebibliography}{10}

\bibitem{Berenstein:2002jq}
D.~E. Berenstein, J.~M. Maldacena, and H.~S. Nastase, {\it {Strings in flat
  space and pp waves from N=4 superYang-Mills}},  {\em JHEP} {\bf 04} (2002)
  013, [\href{http://arxiv.org/abs/hep-th/0202021}{{\tt hep-th/0202021}}].

\bibitem{Banks:1996vh}
T.~Banks, W.~Fischler, S.~H. Shenker, and L.~Susskind, {\it {M theory as a
  matrix model: A Conjecture}},  {\em Phys. Rev. D} {\bf 55} (1997) 5112--5128,
  [\href{http://arxiv.org/abs/hep-th/9610043}{{\tt hep-th/9610043}}].

\bibitem{Kowalski-Glikman:1984qtj}
J.~Kowalski-Glikman, {\it {Vacuum States in Supersymmetric Kaluza-Klein
  Theory}},  {\em Phys. Lett. B} {\bf 134} (1984) 194--196.

\bibitem{Susskind:1997cw}
L.~Susskind, {\it {Another conjecture about M(atrix) theory}},
  \href{http://arxiv.org/abs/hep-th/9704080}{{\tt hep-th/9704080}}.

\bibitem{Sen:1997we}
A.~Sen, {\it {D0-branes on T**n and matrix theory}},  {\em Adv. Theor. Math.
  Phys.} {\bf 2} (1998) 51--59,
  [\href{http://arxiv.org/abs/hep-th/9709220}{{\tt hep-th/9709220}}].

\bibitem{Seiberg:1997ad}
N.~Seiberg, {\it {Why is the matrix model correct?}},  {\em Phys. Rev. Lett.}
  {\bf 79} (1997) 3577--3580, [\href{http://arxiv.org/abs/hep-th/9710009}{{\tt
  hep-th/9710009}}].

\bibitem{Itzhaki:1998dd}
N.~Itzhaki, J.~M. Maldacena, J.~Sonnenschein, and S.~Yankielowicz, {\it
  {Supergravity and the large N limit of theories with sixteen supercharges}},
  {\em Phys. Rev. D} {\bf 58} (1998) 046004,
  [\href{http://arxiv.org/abs/hep-th/9802042}{{\tt hep-th/9802042}}].

\bibitem{Polchinski:1999br}
J.~Polchinski, {\it {M theory and the light cone}},  {\em Prog. Theor. Phys.
  Suppl.} {\bf 134} (1999) 158--170,
  [\href{http://arxiv.org/abs/hep-th/9903165}{{\tt hep-th/9903165}}].

\bibitem{Lin:2004nb}
H.~Lin, O.~Lunin, and J.~M. Maldacena, {\it {Bubbling AdS space and 1/2 BPS
  geometries}},  {\em JHEP} {\bf 10} (2004) 025,
  [\href{http://arxiv.org/abs/hep-th/0409174}{{\tt hep-th/0409174}}].

\bibitem{Lin:2005nh}
H.~Lin and J.~M. Maldacena, {\it {Fivebranes from gauge theory}},  {\em Phys.
  Rev. D} {\bf 74} (2006) 084014,
  [\href{http://arxiv.org/abs/hep-th/0509235}{{\tt hep-th/0509235}}].

\bibitem{Costa:2014wya}
M.~S. Costa, L.~Greenspan, J.~Penedones, and J.~Santos, {\it {Thermodynamics of
  the BMN matrix model at strong coupling}},  {\em JHEP} {\bf 03} (2015) 069,
  [\href{http://arxiv.org/abs/1411.5541}{{\tt arXiv:1411.5541}}].

\bibitem{Witten:1982df}
E.~Witten, {\it {Constraints on Supersymmetry Breaking}},  {\em Nucl. Phys. B}
  {\bf 202} (1982) 253.

\bibitem{Dasgupta:2002hx}
K.~Dasgupta, M.~M. Sheikh-Jabbari, and M.~Van~Raamsdonk, {\it {Matrix
  perturbation theory for M theory on a PP wave}},  {\em JHEP} {\bf 05} (2002)
  056, [\href{http://arxiv.org/abs/hep-th/0205185}{{\tt hep-th/0205185}}].

\bibitem{Kinney:2005ej}
J.~Kinney, J.~M. Maldacena, S.~Minwalla, and S.~Raju, {\it {An Index for 4
  dimensional super conformal theories}},  {\em Commun. Math. Phys.} {\bf 275}
  (2007) 209--254, [\href{http://arxiv.org/abs/hep-th/0510251}{{\tt
  hep-th/0510251}}].

\bibitem{Yi:1997eg}
P.~Yi, {\it {Witten index and threshold bound states of D-branes}},  {\em Nucl.
  Phys. B} {\bf 505} (1997) 307--318,
  [\href{http://arxiv.org/abs/hep-th/9704098}{{\tt hep-th/9704098}}].

\bibitem{Sethi:1997pa}
S.~Sethi and M.~Stern, {\it {D-brane bound states redux}},  {\em Commun. Math.
  Phys.} {\bf 194} (1998) 675--705,
  [\href{http://arxiv.org/abs/hep-th/9705046}{{\tt hep-th/9705046}}].

\bibitem{Aharony:2008ug}
O.~Aharony, O.~Bergman, D.~L. Jafferis, and J.~Maldacena, {\it {N=6
  superconformal Chern-Simons-matter theories, M2-branes and their gravity
  duals}},  {\em JHEP} {\bf 10} (2008) 091,
  [\href{http://arxiv.org/abs/0806.1218}{{\tt arXiv:0806.1218}}].

\bibitem{Kim:2009wb}
S.~Kim, {\it {The Complete superconformal index for N=6 Chern-Simons theory}},
  {\em Nucl. Phys. B} {\bf 821} (2009) 241--284,
  [\href{http://arxiv.org/abs/0903.4172}{{\tt arXiv:0903.4172}}]. [Erratum:
  Nucl.Phys.B 864, 884 (2012)].

\bibitem{Maldacena:2002rb}
J.~M. Maldacena, M.~M. Sheikh-Jabbari, and M.~Van~Raamsdonk, {\it {Transverse
  five-branes in matrix theory}},  {\em JHEP} {\bf 01} (2003) 038,
  [\href{http://arxiv.org/abs/hep-th/0211139}{{\tt hep-th/0211139}}].

\bibitem{Kim:2002if}
N.~Kim and J.~Plefka, {\it {On the spectrum of PP wave matrix theory}},  {\em
  Nucl. Phys. B} {\bf 643} (2002) 31--48,
  [\href{http://arxiv.org/abs/hep-th/0207034}{{\tt hep-th/0207034}}].

\bibitem{Dasgupta:2002ru}
K.~Dasgupta, M.~M. Sheikh-Jabbari, and M.~Van~Raamsdonk, {\it {Protected
  multiplets of M theory on a plane wave}},  {\em JHEP} {\bf 09} (2002) 021,
  [\href{http://arxiv.org/abs/hep-th/0207050}{{\tt hep-th/0207050}}].

\bibitem{Kim:2002zg}
N.~Kim and J.-H. Park, {\it {Superalgebra for M theory on a pp wave}},  {\em
  Phys. Rev. D} {\bf 66} (2002) 106007,
  [\href{http://arxiv.org/abs/hep-th/0207061}{{\tt hep-th/0207061}}].

\bibitem{Choi:2023znd}
S.~Choi, S.~Kim, E.~Lee, S.~Lee, and J.~Park, {\it {Towards quantum black hole
  microstates}},  {\em JHEP} {\bf 11} (2023) 175,
  [\href{http://arxiv.org/abs/2304.10155}{{\tt arXiv:2304.10155}}].

\bibitem{Choi:2023vdm}
J.~Choi, S.~Choi, S.~Kim, J.~Lee, and S.~Lee, {\it {Finite $N$ black hole
  cohomologies}},  \href{http://arxiv.org/abs/2312.16443}{{\tt
  arXiv:2312.16443}}.

\bibitem{Copetti:2020dil}
C.~Copetti, A.~Grassi, Z.~Komargodski, and L.~Tizzano, {\it {Delayed
  deconfinement and the Hawking-Page transition}},  {\em JHEP} {\bf 04} (2022)
  132, [\href{http://arxiv.org/abs/2008.04950}{{\tt arXiv:2008.04950}}].

\bibitem{Choi:2021lbk}
S.~Choi, S.~Jeong, and S.~Kim, {\it {The Yang-Mills duals of small AdS black
  holes}},  \href{http://arxiv.org/abs/2103.01401}{{\tt arXiv:2103.01401}}.

\bibitem{Choi:2018hmj}
S.~Choi, J.~Kim, S.~Kim, and J.~Nahmgoong, {\it {Large AdS black holes from
  QFT}},  \href{http://arxiv.org/abs/1810.12067}{{\tt arXiv:1810.12067}}.

\bibitem{Cabo-Bizet:2018ehj}
A.~Cabo-Bizet, D.~Cassani, D.~Martelli, and S.~Murthy, {\it {Microscopic origin
  of the Bekenstein-Hawking entropy of supersymmetric AdS$_{5}$ black holes}},
  {\em JHEP} {\bf 10} (2019) 062, [\href{http://arxiv.org/abs/1810.11442}{{\tt
  arXiv:1810.11442}}].

\bibitem{Benini:2018ywd}
F.~Benini and E.~Milan, {\it {Black Holes in 4D $\mathcal{N}$=4
  Super-Yang-Mills Field Theory}},  {\em Phys. Rev. X} {\bf 10} (2020), no.~2
  021037, [\href{http://arxiv.org/abs/1812.09613}{{\tt arXiv:1812.09613}}].

\bibitem{Catterall:2010gf}
S.~Catterall and G.~van Anders, {\it {First Results from Lattice Simulation of
  the PWMM}},  {\em JHEP} {\bf 09} (2010) 088,
  [\href{http://arxiv.org/abs/1003.4952}{{\tt arXiv:1003.4952}}].

\bibitem{Asano:2018nol}
Y.~Asano, V.~G. Filev, S.~Kov\'a\v{c}ik, and D.~O'Connor, {\it {The
  non-perturbative phase diagram of the BMN matrix model}},  {\em JHEP} {\bf
  07} (2018) 152, [\href{http://arxiv.org/abs/1805.05314}{{\tt
  arXiv:1805.05314}}].

\bibitem{Schaich:2020ubh}
D.~Schaich, R.~G. Jha, and A.~Joseph, {\it {Thermal phase structure of a
  supersymmetric matrix model}},  {\em PoS} {\bf LATTICE2019} (2020) 069,
  [\href{http://arxiv.org/abs/2003.01298}{{\tt arXiv:2003.01298}}].

\bibitem{Bergner:2021goh}
{\bf Monte Carlo String/M-theory (MCSMC), MCSMC} Collaboration, G.~Bergner,
  N.~Bodendorfer, M.~Hanada, S.~Pateloudis, E.~Rinaldi, A.~Sch\"afer,
  P.~Vranas, and H.~Watanabe, {\it {Confinement/deconfinement transition in the
  D0-brane matrix model \textemdash{} A signature of M-theory?}},  {\em JHEP}
  {\bf 05} (2022) 096, [\href{http://arxiv.org/abs/2110.01312}{{\tt
  arXiv:2110.01312}}].

\bibitem{Schaich:2022duk}
D.~Schaich, R.~G. Jha, and A.~Joseph, {\it {Thermal phase structure of
  dimensionally reduced super-Yang--Mills}},  {\em PoS} {\bf LATTICE2021}
  (2022) 187, [\href{http://arxiv.org/abs/2201.03097}{{\tt arXiv:2201.03097}}].

\bibitem{Pateloudis:2022oos}
S.~Pateloudis, G.~Bergner, N.~Bodendorfer, M.~Hanada, E.~Rinaldi, and
  A.~Sch\"afer, {\it {Nonperturbative test of the Maldacena-Milekhin conjecture
  for the BMN matrix model}},  {\em JHEP} {\bf 08} (2022) 178,
  [\href{http://arxiv.org/abs/2205.06098}{{\tt arXiv:2205.06098}}].

\bibitem{Pateloudis:2022ijr}
{\bf Monte Carlo String/M-theory (MCSMC)} Collaboration, S.~Pateloudis,
  G.~Bergner, M.~Hanada, E.~Rinaldi, A.~Sch\"afer, P.~Vranas, H.~Watanabe, and
  N.~Bodendorfer, {\it {Precision test of gauge/gravity duality in D0-brane
  matrix model at low temperature}},  {\em JHEP} {\bf 03} (2023) 071,
  [\href{http://arxiv.org/abs/2210.04881}{{\tt arXiv:2210.04881}}].

\bibitem{Asano:2012zt}
Y.~Asano, G.~Ishiki, T.~Okada, and S.~Shimasaki, {\it {Exact results for
  perturbative partition functions of theories with SU(2|4) symmetry}},  {\em
  JHEP} {\bf 02} (2013) 148, [\href{http://arxiv.org/abs/1211.0364}{{\tt
  arXiv:1211.0364}}].

\bibitem{Asano:2014vba}
Y.~Asano, G.~Ishiki, T.~Okada, and S.~Shimasaki, {\it {Emergent bubbling
  geometries in the plane wave matrix model}},  {\em JHEP} {\bf 05} (2014) 075,
  [\href{http://arxiv.org/abs/1401.5079}{{\tt arXiv:1401.5079}}].

\bibitem{Asano:2014eca}
Y.~Asano, G.~Ishiki, and S.~Shimasaki, {\it {Emergent bubbling geometries in
  gauge theories with SU(2|4) symmetry}},  {\em JHEP} {\bf 09} (2014) 137,
  [\href{http://arxiv.org/abs/1406.1337}{{\tt arXiv:1406.1337}}].

\bibitem{Asano:2017xiy}
Y.~Asano, G.~Ishiki, S.~Shimasaki, and S.~Terashima, {\it {Spherical transverse
  M5-branes in matrix theory}},  {\em Phys. Rev. D} {\bf 96} (2017), no.~12
  126003, [\href{http://arxiv.org/abs/1701.07140}{{\tt arXiv:1701.07140}}].

\bibitem{Asano:2017nxw}
Y.~Asano, G.~Ishiki, S.~Shimasaki, and S.~Terashima, {\it {Spherical transverse
  M5-branes from the plane wave matrix model}},  {\em JHEP} {\bf 02} (2018)
  076, [\href{http://arxiv.org/abs/1711.07681}{{\tt arXiv:1711.07681}}].

\bibitem{Grant:2008sk}
L.~Grant, P.~A. Grassi, S.~Kim, and S.~Minwalla, {\it {Comments on 1/16 BPS
  Quantum States and Classical Configurations}},  {\em JHEP} {\bf 05} (2008)
  049, [\href{http://arxiv.org/abs/0803.4183}{{\tt arXiv:0803.4183}}].

\bibitem{Chang:2013fba}
C.-M. Chang and X.~Yin, {\it {1/16 BPS states in $\mathcal N=$ 4
  super-Yang-Mills theory}},  {\em Phys. Rev. D} {\bf 88} (2013), no.~10
  106005, [\href{http://arxiv.org/abs/1305.6314}{{\tt arXiv:1305.6314}}].

\bibitem{Chang:2022mjp}
C.-M. Chang and Y.-H. Lin, {\it {Words to describe a black hole}},  {\em JHEP}
  {\bf 02} (2023) 109, [\href{http://arxiv.org/abs/2209.06728}{{\tt
  arXiv:2209.06728}}].

\bibitem{Choi:2022caq}
S.~Choi, S.~Kim, E.~Lee, and J.~Park, {\it {The shape of non-graviton operators
  for $SU(2)$}},  \href{http://arxiv.org/abs/2209.12696}{{\tt
  arXiv:2209.12696}}.

\bibitem{Budzik:2023xbr}
K.~Budzik, D.~Gaiotto, J.~Kulp, B.~R. Williams, J.~Wu, and M.~Yu, {\it
  {Semi-Chiral Operators in 4d ${N}=1$ Gauge Theories}},
  \href{http://arxiv.org/abs/2306.01039}{{\tt arXiv:2306.01039}}.

\bibitem{Chang:2024zqi}
C.-M. Chang and Y.-H. Lin, {\it {Holographic covering and the fortuity of black
  holes}},  \href{http://arxiv.org/abs/2402.10129}{{\tt arXiv:2402.10129}}.

\end{thebibliography}\endgroup
\bibliographystyle{JHEP}

\end{document}